# First experimental evaluation of ambient backscatter communications with massive MIMO reader

R. Fara[1,2], N. Bel-Haj-Maati[2], D.-T. Phan-Huy[2], N. Malhouroux[2], M. Di Renzo[1]

[1]*Laboratoire des Signaux et Systèmes, University of Paris-Saclay, CNRS, CentraleSupélec,* Gif-Sur-Yvette, France
[2]*Orange Labs Networks,* Châtillon & Belfort, France.
romain.fara@orange.com

*Abstract*—Ambient backscatter communications have been introduced as low-power communications for green networking. This technology is very promising as it recycles ambient radio frequency waves, however such systems have limitations and suffer from poor performance due to their low-power. In this paper, we present and evaluate the performance of an ambient backscatter system equipped with a massive multiple input multiple output (MIMO) antenna at the reader side. A mobile device transmits a signal that is backscattered by a tag and received by the reader. Thanks to the spatial diversity of the massive MIMO antenna, the reader is able to identify, for each received signal, the state of the tag (backscattering or transparent) that corresponds to a bit of the tag message. First, we experimentally determine the channel between the device and the reader for two states of the tag. Then using a minimum mean square error algorithm we evaluate the performance of the ambient backscatter communication. We demonstrate that the performance of the ambient backscatter system is significantly improved by the massive MIMO reader.

## I. INTRODUCTION

Internet of Things (IoT) has the potential to provide wireless connectivity between a large number of devices. This important growth in the number of devices is continually increasing energy consumption of wireless networks [1].

In 2013, ambient backscatter (AmB) communications have been introduced as low-power technology [2]. In AmB system, a Radio-Frequency (RF) tag transmits a binary message to a RF reader without battery and without generating additional wave. The tag is illuminated by an RF ambient source of the environment such as: a base station, a TV tower or a Wi-Fi hotspot. The tag transmits "1" and "0" bits by backscattering or not the ambient waves, each bit corresponds to one state of the tag (backscattering and transparent respectively). A tag based on a dipole antenna changes its state by switching the load impedance connected to its antenna. An close to 0 impedance corresponds to the backscattering state: the antenna is short-circuited. An infinite impedance corresponds to a transparent state: the antenna is open-circuited. The state of the tag impacts the ambient waves and can be detected by a simple energy detector reader. As AmB systems do not generate additional wave they are ultra-low power and can be powered by energy-harvesting. Indeed, ambient backscatter communications have been identified as a promising technology for future networks and green communications [3].

However, such technology has limitations, in [4] it was observed that a simple tag and receiver made of one dipole antenna suffers from interference and fading. The tag may not be well illuminated by the source and not able to backscatter signal to the reader. The tag may also interfere with the legacy signal and the reader may not be able to detect the backscattered signal.

Multiple-input multiple-output antennas (MIMO) have largely been studied in the past years, especially for 5G networks. MIMO antennas exploit the spatial diversity of the propagation channel to increase the robustness or the throughput of the wireless communication link. Additionally, it was shown in [5] that 5G MIMO antenna can provide large benefits for low power Internet of Things (IoT) devices. Indeed, the space-time digital processing capability of the massive MIMO 5G base station can be used to detect low power connected objects.

A 5G massive MIMO antenna has been proposed as a smart source for AmB communications in [6]. In downlink configuration, the source beamforms the signal to create a hot spot on the tag and to avoid interference at the reader side. These promising results were obtained through simulations with simple models. In this system the reader was composed of one antenna and was not exploiting the receive diversity to detect the tag.

In [7] and [8], an AmB WiFi system is proposed where source, tag and reader are equipped with MIMO antennas. Such system exploits channel diversity thanks to multiple radiating elements per antenna. However in experiments, the number of receiving antenna elements was limited to a maximum of 4 elements for the source, tag and reader.

MIMO reader is studied in [9] to overcome the limitations of AmB systems. A low complexity algorithm is proposed for the multi-antenna reader to optimize the throughput. However, tag used is not entirely passive and this system has not been experimented and tested in real environment.

In this paper, for the first time, we propose to evaluate experimentally an AmB system using a massive MIMO base station as reader. As illustrated in Fig. 1, we consider an uplink transmission scenario in an anechoic chamber. First, a device, considered as source in the AmB system, sends pilots to the reader. The AmB tag, in the environment, impacts the propagation channel between the device and the reader. The considered tag has two states: a backscattering state in which the dipole antenna is short-circuited (null impedance) and a transparent state in which the antenna is open-circuited (infinite impedance). The propagation channel between the device and the reader depends on the tag state. The reader receives the signal and estimates the propagation channel for the backscattering and the transparent state of the tag. As each state codes a bit of the tag message, the reader is able to decode the tag message.

We propose to exploit the channel spatial diversity using an antenna with 64 elements as reader. The considered massive MIMO antenna is optimized for a carrier frequency between 3.6-3.8 GHz. This antenna has been designed for space-time channel sounding and presented in [10]. The antenna is a Slant Uniform Planar Array (SUPA), composed of 32 elements in vertical polarization and 32 in horizontal polarization with a uniform spacing of half-wavelength.

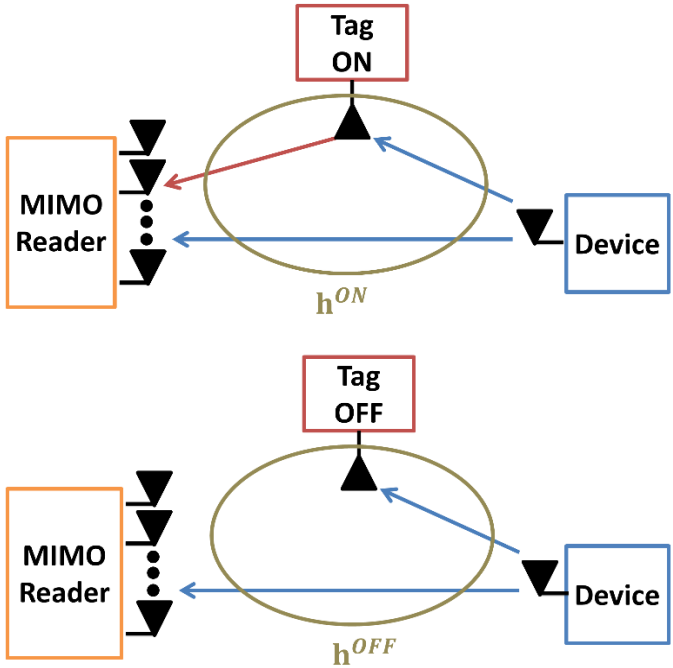

Fig. 1. Backscattering system using MIMO reader in uplink configuration

Based on the experimental channel estimation campaign, we propose to evaluate the performance of the system using a least square error (LSE) estimator to detect the state of the tag. The performance is evaluated by adding additive white gaussian noise (AWGN) to the received signal based on the real channel estimation. Hence, the performance can be evaluated as function of the signal to noise ratio (SNR).

The paper is organized as follows: section II presents our system model, section III presents experimental setup, section IV presents our results and the section IV concludes this paper.

## II. SYSTEM MODEL

### A. Channel estimation

As illustrated in Fig. 1, we consider an AmB system, in an uplink configuration, with a mobile device as a source and a massive MIMO antenna as a reader. The reader is equipped with a SUPA of $K$ elements with uniform spacing of half-wavelength.

We consider a single-antenna tag with two states: a transparent state (OFF) and a backscattering state (ON), depending on the tag state, the channel between the device and the reader is slightly different as shown in Fig. 1.

In order to use a realistic model of the massive MIMO antenna and the propagation channel, the channel impulse response (CIR) is measured for each antenna element of the SUPA. The measured CIR is a discrete channel acquired for a maximum excess delay $T_{max}$, with symbol duration of $T_s$. The channel coefficients are determined for a bandwidth $B = 1/(T_s)$. The CIR is measured $M$ times in order to average the measurement noise. The $m^{th}$ experimental measurement of the CIR $\mathbf{h}^{(m)}$ is defined as:

$$\mathbf{h}^{(m)}(t) = \sum_{q=0}^{Q-1} h^{(m,q)} \delta(t - qT_s), \quad (1)$$

where $h^{(m,q)} \in \mathbb{C}$ is a complex gain, $q$ the time index and $Q$ the number of time index. We have $(Q-1)T_s$, the maximum excess delay.

The estimated channel coefficient vector $\mathbf{h}$ is determined by averaging the $M$ experimental measurement. Therefore $\mathbf{h}$ is defined as:

$$\mathbf{h}(qT_s) = \frac{1}{M} \sum_{m=1}^{M} \mathbf{h}^{(m)}(qT_s), \quad (2)$$

The reader stores $\mathbf{h}$ for the two states of the tag. We denote $\mathbf{h}^{(OFF)}$ and $\mathbf{h}^{(ON)}$, as the stored $\mathbf{h}$ for the transparent and backscattering states, respectively.

The signal received for the $q^{th}$ time index, over the $K$ antenna ports of the reader is $\mathbf{y}(qT_s) \in \mathbb{C}^{K \times 1}$, it is defined for the antenna $k \in [1 \ldots K]$ as:

$$\mathbf{y}_k(qT_s) = \mathbf{h}_k(qT_s)\sqrt{P_u} + \mathbf{v}_k(qT_s), \quad (3)$$

where and $\mathbf{h}_k(qT_s)$ the normalized channel coefficient between the device D and the antenna element $k$ for the bandwidth $B$, $\mathbf{v}_k(qT_s)$ is the noise for the element $k$ of the reader, and $P_u$ is the transmit power spectral density. We denote $P_{noise} = \frac{E[\|\mathbf{v}\|^2]}{K}$, as the received noise power spectral density, where $\|.\|^2$ is the Frobenius norm. We define a signal to noise ratio metric SNR:

$$SNR = \frac{P_u}{P_{noise}}. \quad (4)$$

*B. LSE Estimator*

The reader compares $\mathbf{y}$ and the stored estimation $\mathbf{h}$ and calculated $\hat{y}$ the mean square error:

$$\hat{y} = \arg \min \left\{ \sum_{q=0}^{Q-1} \|\mathbf{z}(qT_s) - \mathbf{y}(qT_s)\|^2 \right\} \quad (5)$$

$with\ \mathbf{z}(qT_s) \in \{\mathbf{h}^{(ON)}(qT_s), \mathbf{h}^{(OFF)}(qT_s)\}.$

Based on the Equation (5) we obtain two values of $\hat{y}$, for $\mathbf{z} = \mathbf{h}^{(ON)}$ and for $\mathbf{z} = \mathbf{h}^{(OFF)}$, the estimator keeps the minimum value between the two. It associates an estimated bit equals to 1 if the backscattering state is detected (ON) and 0 if the transparent state is detected (OFF).

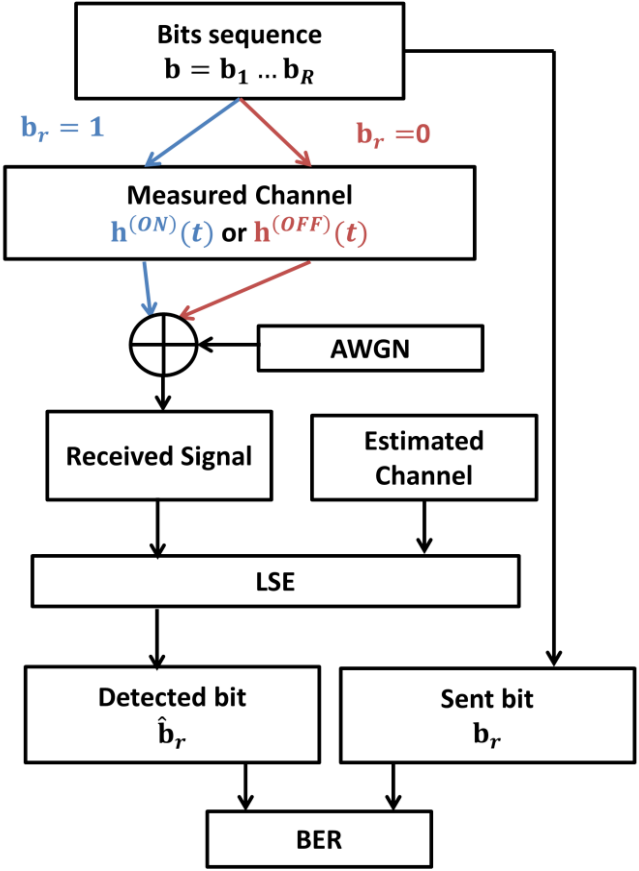

Fig. 2. Overview of system detector

To evaluate the general performance of the system, we consider a random sequence of $R$ sent bits $\mathbf{b} = \mathbf{b}_1 \ldots \mathbf{b}_r \ldots \mathbf{b}_R$, if $\mathbf{b}_r = 1$ (or $\mathbf{0}$ respectively) we use $\mathbf{h} = \mathbf{h}^{(ON)}$ (or $\mathbf{h}^{(OFF)}$ respectively) in Equation (3). Then, we deduce a sequence of $R$ bits from the LSE detector $\hat{\mathbf{b}} = \hat{\mathbf{b}}_1 \ldots \hat{\mathbf{b}}_r \ldots \hat{\mathbf{b}}_R$, we compare the detected bits $\hat{\mathbf{b}}$ with the sent bits $\mathbf{b}$ and we calculate the *BER* as follows:

$$BER = \frac{\sum_{r=1}^{R} |\hat{\mathbf{b}}_r - \mathbf{b}_r|^2}{R}. \quad (6)$$

III. EXPERIMENTAL SETUP

CIR measurements were conducted in an anechoic chamber. The experimental system consisted of a 64 elements massive MIMO antenna, a device to illuminate a two states tag. These elements are illustrated in the complete system in Fig. 3 and in Fig. 4 as separated elements.

For a sample acquisition, the device, illustrated in Fig. 4-c), transmits a pulse signal at the carrier frequency of 3.7 GHz. The massive MIMO antenna, illustrated in Fig. 4-a), receives the device signal through the propagation channel acquires the signal on each of the 64 elements. The system response is measured for a maximum excess delay of $4.10^{-8}$ seconds for different antenna and bandwidth configurations.

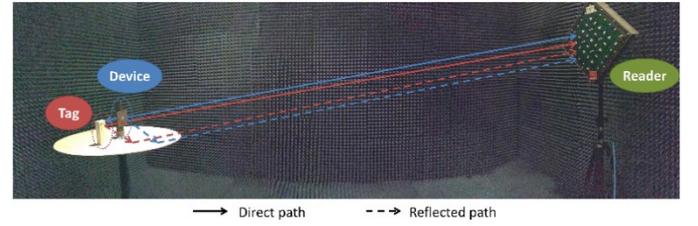

Fig. 3. Photo of the complete experimental system

The two states tag is illustrated in Fig. 4-b), we observe the half wavelength dipole antennas and a switch that allows us to change the state of the tag. The switch can open-circuit the antenna branch (transparent) and short-circuit the branch of the antenna (backscattering).

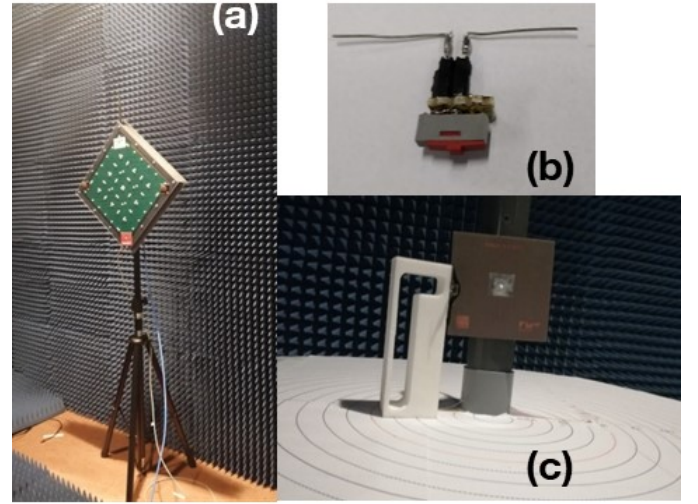

Fig. 4. Photos of the 64 elements antenna (a), of the tag (b), of device and tag (c)

The channel has been measured for a distance separating the tag and the device equal to $2\lambda$, and for a fixed position of the device, the tag and the MIMO antenna. The measurement has been done for the two states of the tag (transparent and backscattering). Each channel acquisition was repeated between $M=5$ to $M=6$ times.

TABLE I. EXPERIMENTAL PARAMETERS

| Parameters | Details | Value | Units |
|---|---|---|---|
| $K$ | Number of Reader Antenna | {1, 2, 4, 8, 16, 32, 64} | |
| $R$ | Number of bits per sequence | $10^6$ | |
| $d^{D-T}$ | Distance between the Device and the tag | $2\lambda$ | m |
| $M^{ON}$ | Number of acquisition (ON state) | 6 | |
| $M^{OFF}$ | Number of acquisition (OFF state) | 5 | |

The gains and the phases for each elements of the MIMO antenna are obtained with the measurement campaign. Based on these acquisitions, the 64 channels are evaluated for the two states of the tag.

In this simulation we evaluate the performance of the tag state detection using a LSE algorithm for a SNR between -10 dB and 20 dB.

### A. Massive MIMO Antenna

We evaluate the performance as function of the number of receiving antenna elements. The different configurations are illustrated in Fig.5. The number K of antenna elements depends on the number of active elements on the reader antenna. The configurations are chosen in such a way that the total spatial occupancy of the active elements increases with K. This enables us to emulate a large variety of readers: from small devices, laptops, tablets, small cells, MIMO base stations, up to massive MIMO base stations.

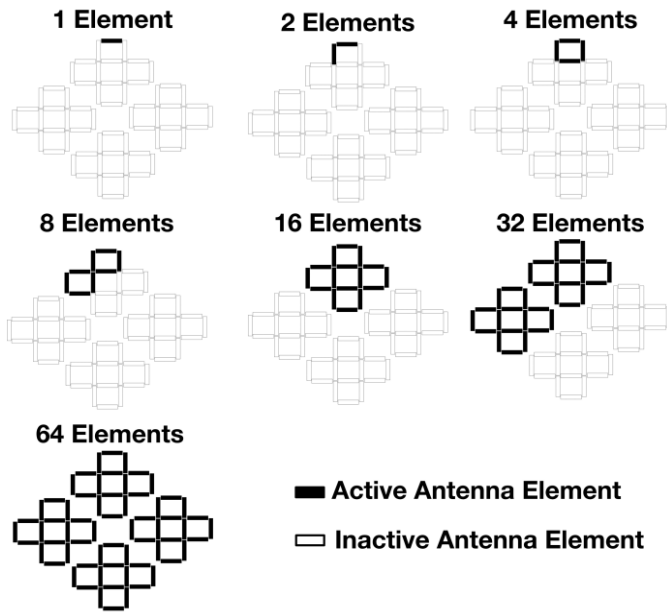

Fig. 5. Configurations of the MIMO antenna.

### B. Bandwidth

The performance of the system is also evaluated as a function of the bandwidth. We consider 3 different configurations for the bandwidth of the measured signal, detailed in Table II. For each configuration, we keep the same central frequency, as illustrated in Fig. 6.

TABLE II. BANDWIDTH CONFIGURATIONS

| Bandwidth Configuration | B1 | B2 | B3 |
|---|---|---|---|
| Bandwidth (MHz) : B | 100 | 60 | 20 |
| Number of time samples per acquisition : $Q$ | 5 | 3 | 1 |
| Symbol duration (seconds) : $T_s$ | $1.10^{-8}$ | $2.10^{-8}$ | $5.10^{-8}$ |
| Maximum excess delay (seconds) : $T_{max}$ | $40.10^{-8}$ | $40.10^{-8}$ | $40.10^{-8}$ |

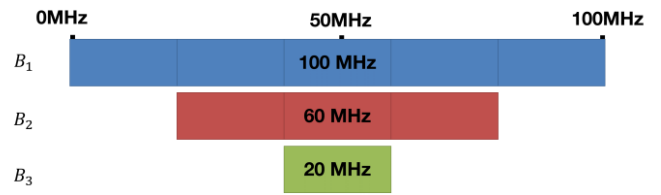

Fig. 6. Configurations of the bandwidth

In order to allow fair comparison between the three configurations, the power spectral densities of the signal and the noise are kept identical when changing the bandwidth. As a consequence, when we consider a large bandwidth, the number of symbols increases but the SNR (per symbol) is the same.

## IV. RESULTS

This section presents experimental results based on the model depicted in II, and the setup given in Section III.

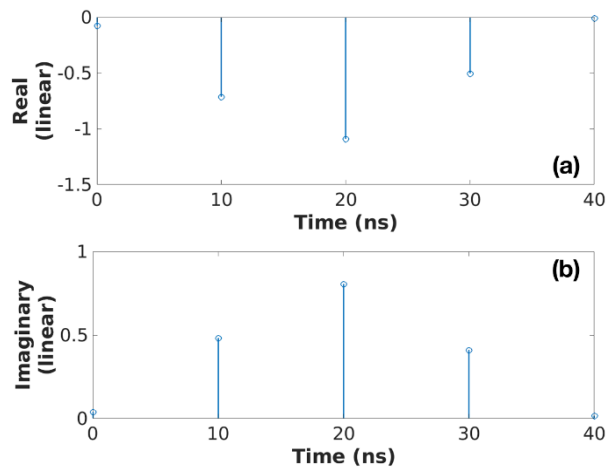

Fig. 7. Exemple of CIR as function of the time for the first antenna element : a) real part b) imaginary part

We first observe in Fig. 7 the amplitude of the experimental CIR of the system for the first element of the reader antenna as function of the time. The channel has several echoes corresponding to the tag and the support table. The support creates a non-negligible reflection even if the material of the support is not very sensitive to electromagnetic waves.

In Fig. 8, the 64 elements of the massive MIMO antenna are positioned on a grid of 8 lines and 8 columns. The maps illustrate the spatial signature in line as function of the time for the two states. The top maps (a) illustrate the amplitude of the CIR coefficients for $qT_s = 0$ to 40 ns. The bottom maps (b) illustrate the phase between -180° and 180° of the CIR. We observe the spatial diversity of the massive MIMO antenna. One can note that between the two states of the tag the differences are visible in the phase domain. The amplitude of the signal received for each element of the reader is not significantly impacted by the state of the tag. However, the impact is more significant for the phase domain.

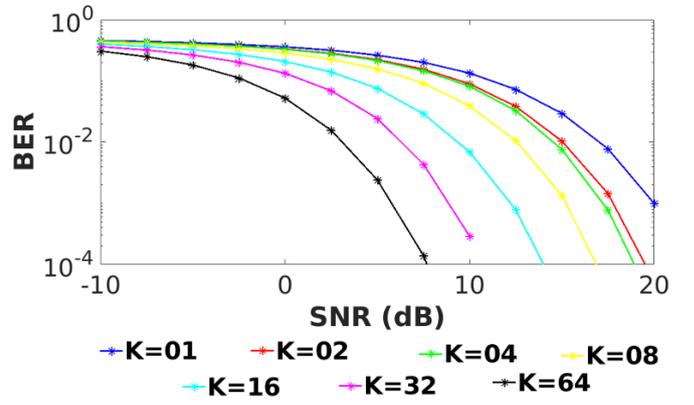

Fig. 9. Bit error rate as function of SNR for different $K$ antenna elements configurations.

Fig. 9 illustrates the BER as function of the SNR for different configurations of $K$ antenna elements. Each antenna configuration is evaluated for the bandwidth configuration B1 (100 MHz). We observe that as the number of antennas increases the performance is improved. The MIMO antenna exploits the channel diversity and this improves the reliability and the robustness of the detection.

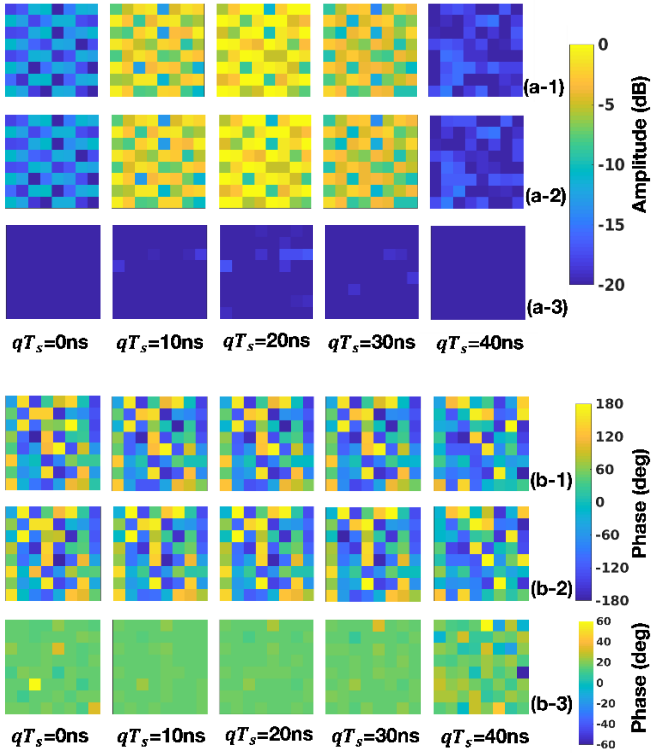

Fig. 8. Exemple of CIRs for the 64 receive antenna elements configurations. a-1) amplitude of the CIR for the transparent state, a-2) amplitude of the CIR for the backscattering state, a-3) amplitude difference between the two states b-1) phase of the CIR for the transparent state, b-2) phase of the CIR for the backscattering state, b-3) difference of phase between the two states (on average equal to 16 degrees).

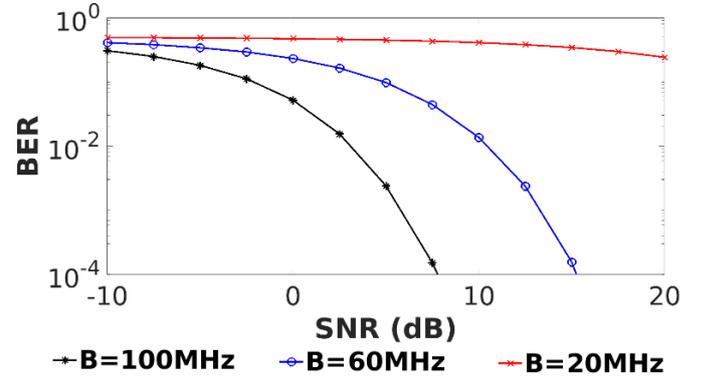

Fig. 10. Bit error rate as function of SNR for different bandwidth configurations.

Fig. 10 illustrates the BER as function of the SNR for different bandwidth configurations. Each bandwidth configuration is evaluated for $K = 64$ antenna elements. We observe that the performance is improved if the bandwidth increases. This is due to the fact that the frequency diversity of the channel increases with the bandwidth. Indeed, the impact of the echoes, which create frequency diversity, increases with the bandwidth. Diversity improves the performance of the LSE detector. The performance of the bandwidth $B = 20$ MHz is poor because the bandwidth is too reduced to have the impact of the echoes.

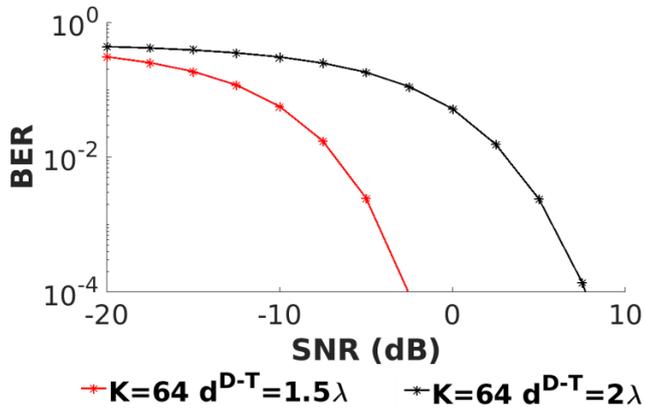

Fig. 11. Bit error rate as function of SNR for different device to tag distance configurations.

Finally, we evaluate the performance depending on the distance between the device and the tag. Device location is fixed and the tag is located at 1.5 or 2 lambda of the device. In the Fig. 11 we observe that the performance is impacted by the distance. This effect has been observed in SISO with ellipses in [4] and was due to the phase of the signal. As we observe in Fig. 12 the phase difference is more significant for the configuration $d^{D-T} = 1.5\lambda$ and may result in the performance improvement. For further studies, remains the performance evaluation with respect to the distance between the tag and the device in a MIMO reader configuration.

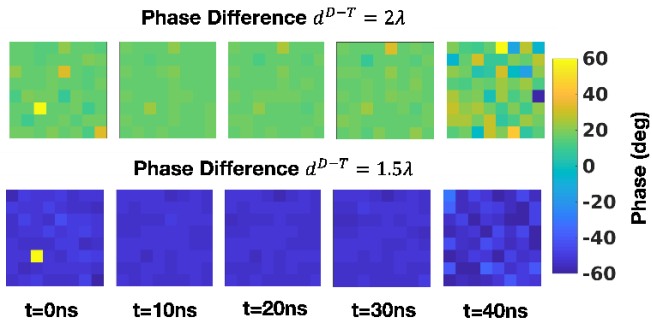

Fig. 12. Phase difference of the CIR between the two states (transparent and backscattering) for different distance device to tag configurations. 64 elements of the massive MIMO antenna are positioned on a grid of 8 lines and 8 columns.

## V. CONCLUSION

In this paper, for the first time, we have evaluated the performance of an ambient backscatter system with a massive multiple input multiple output antenna as a reader and a mobile device as a source, experimentally. A device sends a signal to the reader that is backscattered or not, by a tag, depending on the tag state. The reader estimates the channel for the two states of the tag (transparent and backscattering) and applies a least square error algorithm to detect the bit sent by the tag corresponding to the state of the tag. This experimental evaluation has been done for multiple values of signal to noise ratio. We show that the spatial diversity of the reader antenna highly improves the performance of the detection. We have also shown the impact of the number of receiving antenna and the bandwidth. Future studies will assess the performance on more complex propagation channels, outside anechoic chamber and multi-state tags.

ACKNOWLEDGMENT

This work is partially supported by the French Project ANR Spatial Modulation under grant ANR-15-CE25-0016 (https://spatialmodulation.eurestools.eu/).

REFERENCES

[1] A. Gati *et. al*., "Key technologies to accelerate the ict green evolution: An operators point of view," Submitted to IEEE Communications Surveys & Tutorials, 2019. Available at: https://arxiv.org/abs/1903.09627.

[2] V. Liu, A. Parks., V. Talla, S. Gollakota, D. Wetherall, and J. R. Smith, "Ambient backscatter: Wireless communication out of thin air," in *Proc. SIGCOMM 2013,* 2013.

[3] W. Zhang, Y. Qin, W. Zhao, M. Jia, Q. Liu, R. He, B. Ai, "A green paradigm for Internet of Things: Ambient backscatter communications," *China Communications*, vol. 16, no. 7, pp. 109-119, 2019.

[4] K. Rachedi, D.-T. Phan-Huy, N. Selmene, A. Ourir, M. Gautier, A. Gati, A. Galindo-Serrano, R. Fara, J. de Rosny "Demo Abstract : Real-Time Ambient Backscatter Demonstration," in *Proc. IEEE INFOCOM 2019*, pp. 987–988, 2019.

[5] D.-T. Phan-Huy *et. al.*, "Single-Carrier Spatial Modulation for the Internetof Things: Design and Performance Evaluation byUsing Real Compact and Reconfigurable Antennas" *IEEE Access*, vol. 7, pp. 18978-18993, 2019.

[6] R. Fara, D.-T. Phan-Huy and M. Di Renzo, "Ambient backscatters-friendly 5G networks: creating hot spots for tags and good spots for readers," Submitted to *IEEE WCNC 2020*.

[7] J. Zhao, W. Gong and J. Liu, "Spatial Stream Backscatter Using Commodity WiFi," in *Proc. 16th Annual International Conference on Mobile Systems, Applications, and Services (MobiSys '18)*, pp 191-203, 2018.

[8] X. Liu, Z. Chi, W. Wang, Y. Yao and T. Zhu, " VMscatter: A Versatile MIMO Backscatter" in Proc *17th USENIX Symposium on Networked Systems Design and Implementation*, pp.895-909, 2020.

[9] D. Mishra and E.G. Larsson, "Multi-Tag Backscattering to MIMO Reader: Channel Estimation and Throughput Fairness," *IEEE Transactions on Wireless Communications,* 18, 5584-5599, 2019.

[10] P. Pajusco, F. Gallée, N. Malhouroux-Gaffet, R. Burghelea. "Massive antenna array for space time channel sounding," in *Proc. EUCAP 2017 : 11th European conference on antennas and propagation*, pp.865 – 868, 2017.